\documentstyle[aps,epsf]{revtex}

\newcommand{\dd}{\text{d}}
\newcommand{\ee}{\text{e}}
\newcommand{\ii}{\text{i}}
\newcommand{\ji}{\text{\footnotesize i}}

\begin{document}
\title{{\bf Repeated bond traversal probabilities}\\
for the simple random walk}
\author{T. Antal$^1$,\, H.J. Hilhorst$^2$,\, and\, R.K.P. Zia$^3$}
\address{{\small $^1$D\'epartement de Physique Th\'eorique,}\\
{\small Universit\'e de Gen\`eve, CH 1211 Gen\`eve 4, Switzerland}\\
{\small $^2$Laboratoire de Physique Th\'eorique, B\^atiment 210,}\\
{\small Universit\'e de Paris-Sud, 91405 Orsay cedex, France}\\
{\small $^3$Center for Stochastic Processes in Science and Engineering,}\\
{\small Department of Physics,}\\
{\small Virginia Polytechnic Institute and State University,}\\
{\small Blacksburg, Virginia 24061, USA}}
\maketitle

\begin{abstract}
\noindent We consider the average number $B_m(t)$ of bonds traversed exactly 
$m$ times by a $t$ step simple random walk. We determine $B_m(t)$ explicitly
in the scaling limit $t\to\infty$ with $m/\sqrt{t}$ fixed in dimension $d=1$
and $\,m/\log t\,$ fixed in dimension $d=2$. The scaling function is an\,
erfc\, in $d=1$ and an exponential in $d=2$.\newline
\end{abstract}


\section{Introduction}

\label{secintroduction}\noindent The simple random walk is a venerable
problem, finding its applications in many areas of statistical physics.
Despite its long history, novel aspects continue to surface. Motivated by
recent experiments on gas transport properties in polycarbonate films, in
which polymers are found in a glassy state, we are led to the the following
question. Suppose the polymers are modeled by the usual (non-interacting)
random walks on a lattice, how many ``monomers'' can we expect on each bond?
Clearly, the answer will depend on both the length and the density of the
polymers. Not surprisingly, it is a simple step once we know the
distribution of bond-traverses of a {\em single} polymer (or random walker).
Remarkably, this distribution is not, to the best of our knowledge, in the
literature. This note will be devoted to our findings of such a distribution.

Let a simple random walk start at the origin of a $d$-dimensional hypercubic
lattice. We wish to know the average number of bonds $B_m(t)$ that it has
traversed exactly $m$ times. This quantity satisfies 
\begin{equation}
\sum_{m=1}^\infty B_m(t)=B(t),\qquad \qquad \sum_{m=1}^\infty mB_m(t)=t
\label{sumrules}
\end{equation}
where $B(t)$ is the average total number of distinct bonds traversed by
(also known as the support) the walk.

If for $t\to\infty$ asymptotically $B(t)\simeq\beta(t)$, then one may
reasonably expect that in that limit $B_m(t)$ can be expressed as a scaling
function of $m\beta(t)/t$. The sum rules (\ref{sumrules}) then imply that $%
B_m(t)$ takes the form 
\begin{equation}
B_m(t)\simeq\frac{\beta^2(t)}{t}{\cal B}\Big(\frac{m\beta(t)}{t}\Big)
\label{scaling}
\end{equation}
where ${\cal B}$ is a scaling function.

In this note 
we will explicit find ${\cal B}$ (and $\beta$) in spatial dimensions $d=1$
and $d=2$, thereby justifying the scaling form Eq.\,(\ref{scaling}). It
appears that ${\cal B}$ is an error function ``erfc'' in $d=1$ and an
exponential in $d=2$. In $d=2$ we determine, moreover, the leading order
correction to the scaling behavior (\ref{scaling}).\newline

\section{Generating function method for repeated bond traversals}

\label{secgenerating} \vspace{4mm}

In order to determine $B_m(t)$ we need the following notation. Let the
vector $\delta$ denote any of the $d$ basis vectors of the lattice. Let $%
(x,\delta)$ denote the bond between the sites $x$ and $x+\delta$ (the fact
that this bond is identical to $(x+\delta,-\delta)$ is of no importance).
Let $B_m(x,\delta;t)$ be the probability that $(x,\delta)$ is traversed
exactly $m$ times, irrespective of the direction. Then 
\begin{equation}
B_m(t) = \sum_{(x,\delta)} B_m(x,\delta;t)  \label{sumonx}
\end{equation}
where each bond occurs exactly once in the summation. The quantity $%
B_m(x,\delta;t)$ may be calculated by an adaptation of the analogous method
for multiple visits to the same {\it site} \cite{Hughes}. Let $F(x,\delta;t)$
be the probability that the first traversal of $(x,\delta)$ occurs at the $t$%
\,th step, with $t=1,2,\ldots\,$ It is convenient to set $%
F(x,\delta;0)\equiv 0$.

For any $X(t)$ defined for $t=0,1,2,\ldots$ we may introduce the generating
function $\hat{X}(z) \equiv \sum_{t=0}^\infty z^t X(t).$ Let furthermore $%
R(\tau)$ be the probability that, given a traversal has taken place, the
next one occurs exactly $\tau$ steps later, with $\tau=1,2,\ldots\,$ Then $%
R(\tau)=F(0,\delta;\tau)$ and hence $\hat{R}(z) = \hat{F}(0,\delta;z).$
Reasoning in a similar way as for the site problem one finds that 
\begin{equation}
\hat{B}_m(z) = \frac{1}{1-z}\,\,\Big[ \sum_{(x,\delta)} \hat{F}(x,\delta;z)%
\Big] \,[1-\hat{R}(z)]\,\hat{R}^{m-1}(z) \qquad (m=1,2,\ldots)  \label{relBF}
\end{equation}
Here the function $\hat{F}$, which implies $\hat{R}$, still have to be found.

Let $G(x;t)$ be the probability that after $t$ steps the walk is at lattice
site $x$, for $t=0,1,2,\ldots \,$ Let $G(x,\delta ;t)$ denote the
probability that at its $t$\thinspace th step it traverses (in either
direction) the bond $(x,\delta )$, for $t=1,2,\ldots \,$ We set additionally 
$G(x,\delta ;0)\equiv 0$. Since $(x,\delta )$ can be traversed starting
either from $x$ or from $x+\delta $, we have 
\begin{equation}
G(x,\delta ;t)=\frac 1{2d}[G(x;t-1)+G(x+\delta ;t-1)]\qquad (t=1,2,\ldots )
\label{relGGt}
\end{equation}
By a slight extension of the standard procedure for calculating first
passage probabilities on sites \cite{Weiss,Hughes,WH} we have here 
\begin{equation}
G(x,\delta ;t)=F(x,\delta ;t)\,+\,\sum_{\tau =0}^tF(x,\delta ;\tau
)G(0,\delta ;t-\tau )\qquad (t=1,2,\ldots )  \label{relFGt}
\end{equation}
In terms of generating functions, Eqs.\thinspace (\ref{relGGt}) and (\ref
{relFGt}) become, respectively, $\hat{G}(x,\delta ;z)=(z/2d)[\hat{G}(x;z)+%
\hat{G}(x+\delta ;z)]$ and $\hat{F}(x,\delta ;z)=\hat{G}(x,\delta ;z)/[1+%
\hat{G}(0,\delta ;z)].$ 
Elimination of $\hat{G}(x,\delta ;z)$ from this pair of equations yields 
\begin{equation}
\hat{F}(x,\delta ;z)=\frac z{2d}\,\,\,\frac{\hat{G}(x;z)+\hat{G}(x+\delta ;z)%
}{1\,+\,\frac z{2d}[\hat{G}(0;z)+\hat{G}(\delta ;z)]}  \label{relFGzfinal}
\end{equation}
This achieves the reduction of the desired function $\hat{F}(x,\delta ;z)$
to the known function $\hat{G}(x;z)$. Note that expression (\ref{relFGzfinal}%
) has the required invariance under the replacement $(x,\delta )\to
(x+\delta ,-\delta )$. We now exploit the well-known relations 
$\hat{G}(\delta ;z)=[\hat{G}(0;z)-1]/z$ and $\sum_x\hat{G}(x;z)=1/(1-z).$
Upon combining these with Eqs.\thinspace (\ref{relBF}) and (\ref{relFGzfinal}%
) one gets, writing henceforth $G(z)\equiv \hat{G}(0;z)$, a fully explicit
expression for the generating function $\hat{B}_m(z)$ in terms of $G(z)$. 
\begin{equation}
\hat{B}_m(z)=\frac z{(1-z)^2}\frac{(2d)^2}{[2d+z\tilde{G}(z)]^2}\Big[\frac{z%
\tilde{G}(z)}{2d+z\tilde{G}(z)}\Big]^{m-1},  \label{Bmz}
\end{equation}
where $\tilde{G}\left( z\right) $ is defined by 
\[
\tilde{G}\equiv z^{-1}\left[ \left( 1+z\right) G\left( z\right) -1\right] . 
\]
The $m$-th coefficient is extracted as 
\begin{equation}
B_m(t)=\frac 1{2\pi \ii}\oint \frac{\dd z}{z^{t+1}}\hat{B}_m(z),
\label{inverse}
\end{equation}
where the integral runs counterclockwise around the origin. To make $B_m(t)$
more explicit, we have to consider each spatial dimension separately.

\section{One dimension}

\label{seconedimension}

In dimension $d=1$ we have to evaluate Eq.\,(\ref{inverse}) with \cite
{Weiss,Hughes} $G(z) = (1-z^2)^{-1/2}.$ It turns out to be advantageous to
consider the differences $B_{m+1}(t)-B_m(t)$. After slight rewriting this
yields 
\begin{equation}
B_{m+1}(t)-B_m(t) = -\frac{1}{\pi\ii}\oint\frac{\dd z}{z^{t+m+2}} \,\,\Big[%
\sqrt{\frac{1+z}{1-z}}-1\Big]\,\Big[1-\sqrt{1-z^2}\Big]^m
\label{intdifference}
\end{equation}
We will now fold the integration path around the branch cut that runs from $%
z=1$ to $z=\infty$ along the positive real axis. In the limit $t\to\infty$
we expect a meaningful result only if also $m\to\infty$ at fixed ratio $m/%
\sqrt{t}$. We anticipate that in this limit the integral on $z$ draws its
main contribution from a region at a distance of order $t^{-1/2}$ from the
branch point at $z=1$. We therefore introduce the scaling variable 
\begin{equation}
\mu_1 = m/\sqrt{2t}  \label{defmu1}
\end{equation}
and the scaled variable of integration $y=(z-1)t.$ The integrand of Eq.\,(%
\ref{intdifference}) may now be expanded in powers of $t^{-1/2}$ at fixed $%
\mu$ and $y$. This leads to 
\begin{equation}
B_{m+1}(t)-B_m(t) = -\,\sqrt{\frac{2}{\pi^2t}}\int_0^\infty\, \frac{\dd y}{%
\sqrt{y}}\, (\ee^{2\ji\mu_1\sqrt{y}} + \ee^{-2\ji\mu_1\sqrt{y}})\,\ee^{-y}
\label{intcut}
\end{equation}
where the two terms on the right hand side come from above and below the
branch cut, respectively. The integral is easily found to be equal to $%
-\,2^{3/2}\,(\pi t)^{-1/2}\,\ee^{-\mu_1^2}$. With the boundary condition $%
B_\infty(t)=0$ we therefore find upon integrating 
\begin{equation}
B_m(t) = 2\,\mbox{erfc}(\mu_1)\,+\,{\cal O}(t^{-1/2})  \label{final1}
\end{equation}
which is the final result, valid for $t\to\infty$ at $\mu_1$ fixed. 
When summing Eq.\,(\ref{final1}) on $m$ (or alternatively when evaluating
the integral obtained by summing Eq.\,(\ref{inverse}) on $m$), one obtains $%
B(t)\simeq\sqrt{8t/\pi}$. Together with Eq.\,(\ref{final1}) 
this confirms the validity of the hypothesized scaling form (\ref{scaling}).

In order to compare the large time scaling behavior of $B_m(t)$ to its
finite time forms we have performed Monte-Carlo simulations. For
$t=10^2$, $10^3$, $10^4$ the bond distribution was averaged over
several independent runs.  Fig.\,1 shows that for increasing values of
$t$ the simulation data rapidly converge to the scaling function, and
they are practically indistinguishable on this figure for $t \ge 10^3$.

\begin{figure}[htb]
  \centerline{ \epsfxsize=9cm \epsfbox{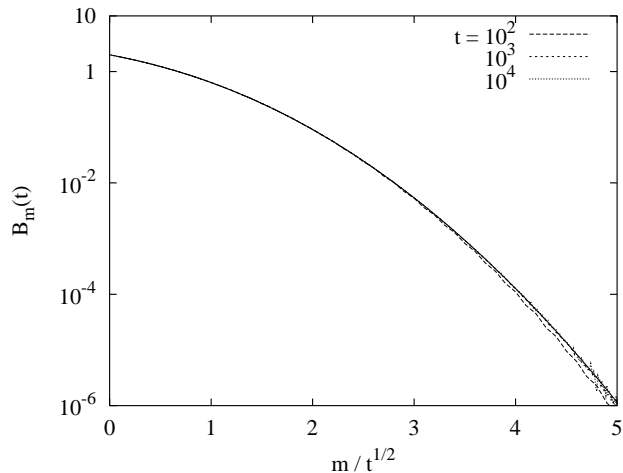} } \vspace{0.5cm}
\caption{Simulation results for $B_m(t)$ in one dimension at different
finite values of time and its scaling form in the $t\to \infty$ limit
(solid line).}
\end{figure}

\section{Two dimensions}

\label{sectwodimensions}

For a two-dimensional square lattice the expansion \cite{Weiss,Hughes} $G(z)
= \frac{1}{\pi}\,\log\frac{8}{1-z}\,+\,{\cal O}((1-z)\log(1-z))$ enables one
to evaluate Eq.\,(\ref{inverse}) for $t$ asymptotically large. The only
integral whose asymptotic behavior we need is \cite{HenyeySeshadri,WH} 
\begin{equation}
\frac{1}{2\pi\ii}\oint\frac{\dd z}{z^{t+1}}\frac{1}{(1-z)^2}\frac{\pi^n}{%
\log^n\frac{8}{1-z}} \,=\, \frac{\pi^n t}{\log^n 8t}\,\Big[ \,1\,+\,\frac{%
n\,(1-C)}{\log 8t} \,+\,{\cal O}(\log^{-2}8t)\Big]  \label{integral}
\end{equation}
where $C=0.577215...$ is Euler's constant. We anticipate a meaningful result
for $t\to\infty$ if also $m\to\infty$ in such a way that the ratio $m/\log t$
remains fixed. We expect that in this scaling limit the $z$ integral will
draw its main contribution from a region in the complex plane near the
branch point $z=1$ where the ratio $m/G(z)$ is finite. Keeping this in mind
we expand the integral (\ref{inverse}) for $B_m(t)$ as 
\begin{equation}
B_m(t) = \frac{1}{2\pi\ii}\oint\frac{\dd z}{z^{t+1}}\frac{1}{(1-z)^2 G^2(z)}
\,\ee^{-2m/G(z)}\,\Big[\,1\,+\,\Big(\frac{m}{G(z)}-1\Big)\frac{1}{G(z)}\,+\, 
{\cal O}(G^{-2}(z))\Big]  \label{intexpansion}
\end{equation}
We may now insert in (\ref{intexpansion}) the explicit expression $G(z)\simeq%
\frac{1}{\pi}\log\frac{8}{1-z}$, knowing that the ${\cal O}((1-z)\log(1-z))$
terms in the expansion of $G(z)$ will contribute only terms with higher
powers of $t^{-1}$ to the final asymptotic series. The exponential in Eq.\,(%
\ref{intexpansion}) may then be expanded, the resulting series integrated
term by term with the aid of Eq.\,(\ref{integral}), and summed again. Upon
introducing the scaling variable 
\begin{equation}
\mu_2 = \frac{2\pi m}{\log 8t}  \label{defmu}
\end{equation}
\vspace{-4mm}

\noindent one finds the final result 
\begin{equation}
B_m(t)=\frac{4\pi ^2t}{\log ^28t}\,\ee^{-\mu _2}\,\Big[\,1\,+\,(\mu _2-2)%
\frac{\frac \pi 2+1-C}{\log 8t}\,+\,{\cal O}(\log ^{-2}8t)\Big]
\label{final2}
\end{equation}
valid for $t\to \infty $ at $\mu _2$ fixed. As expected, the correction
terms decay only as powers of the inverse logarithm of the number of steps.
The next few higher order terms in the asymptotic series (\ref{final2}) may
be calculated without great effort. Since the $k$th order correction term is
multiplied by a $k$th degree polynomial in $\mu _2$, this is actually an
expansion in powers of $\mu _2/\log 8t$. 
One may verify that Eq.\thinspace (\ref{final2}) satisfies the sum rules (%
\ref{sumrules}) up to and including the first order correction term. The
scaling function ${\cal B}$ introduced in Eq.\thinspace (\ref{scaling}) here
appears to be a simple exponential. One finds that the average total number
of bonds traversed increases as $\beta (t)=2\pi t/\log 8t\,+\,{\cal O}(t\log
^{-2}8t)$.

\begin{figure}[htb]
  \centerline{ \epsfxsize=9cm \epsfbox{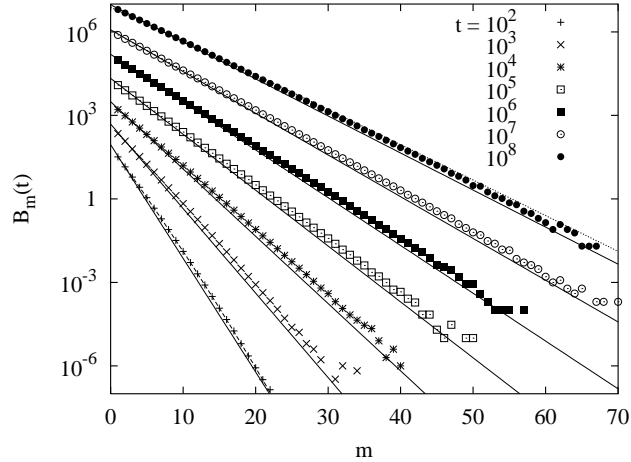} } \vspace{0.5cm}
\caption{Simulation data for $B_m(t)$ in two dimensions at different
values of time (symbols) compared to the theoretical scaling form
(solid lines). The exact result for $t=10^2$ (dashed line) and the
scaling law plus leading correction for $t=10^8$ (dotted line) are
also displayed.}
\end{figure}

Fig.\,2 compares simulation data for $t=10^2, 10^3,\ldots,10^8$ to the
theoretical scaling law (the first term of Eq.\,(\ref{final2})). The
$t=10^2$ data are also compared to the exact curve, which 
we obtained by Taylor expanding Eq.\,(\ref{Bmz}) 
through the one hundredth term with the aid of a symbol
manipulation program. For
$t=10^8$ the scaling law plus leading correction (the first two terms
of Eq.\,(\ref{final2})) is also displayed. Upon collapsing the data of
Fig.\, 2 one finds Fig.\, 3, which shows that in $d=2$ the convergence
to the asymptotic scaling law is very slow.

\begin{figure}[htb]
  \centerline{ \epsfxsize=9cm \epsfbox{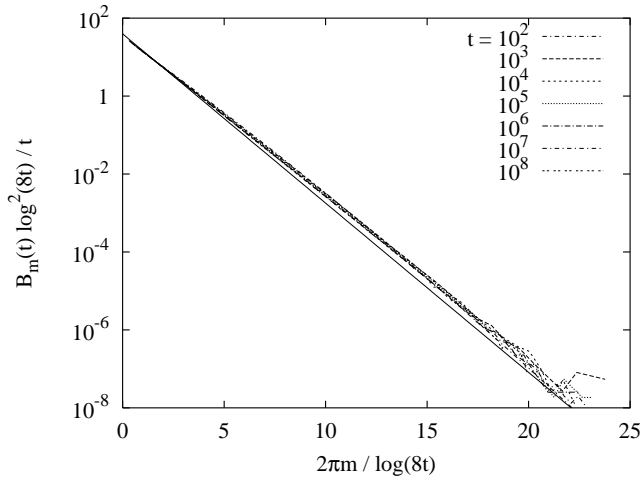} } \vspace{0.5cm}
\caption{Data collapse of two dimensional simulation results for
$B_m(t)$. The $t \to \infty$ scaling limit is shown as a solid
line.}
\end{figure}

\section{Concluding Remarks}

\label{seccomments}We conclude this note with a series of comments and
remarks.

An alternative route to $B_m(t)$ is to consider a complementary process, by
focusing on a {\em particular} bond in, say, a finite periodic $L^d$
lattice. Now, we can ask how often this bond is traversed by a $t$-step
random walk which starts at $x_0$ and ends at $x$. By summing
over $(x,x_0)$ and invoking translational invariance, we
obviously access $B_m(t)$. In this approach, we can easily generalize the
problem to a study of the frequency of traversing in only {\em one
direction. }To solve both problems, consider a modified random walker, for
which the rate of traversing our chosen bond is $p/2d$ instead of $1/2d$.
Solving the master equation for $P(x,t|x_0,0;p)$ (the usual
probability) by standard generating function techniques and defining $\tilde{%
B}\left( p,z\right) \equiv \sum_{x,x_0}\sum_{t=0}^\infty z^tP(%
x,t|x_0,0;p)$, we obtain 
\begin{equation}
\tilde{B}\left( p,z\right) =\frac{L^d}{1-z}-\frac{2z\left( 1-p\right) }{%
\left( 1-z\right) ^2\left[ 2d+\left( 1-p\right) z\tilde{G}\left( z\right)
\right] }\,\,.  \label{Btilde}
\end{equation}
The first term represents, since $\sum_{x}P(x,t|x%
_0,0;p=1)=1$, the $L^d$ points of origination of the walks we considered.
Due to this difference in normalization, the bivariate
generating function
for $B_m(t)$, {\it i.e.}
$\sum_{m=1}^\infty \sum_{t=0}^\infty p^mz^tB_m(t)$, is
precisely $d\left[ \tilde{B}\left( p,z\right) -\tilde{B}\left( 0,z\right)
\right] .$ With the extra factor $d$, 
the quantity $d\tilde{B}\left( 1,z\right) $
represents nothing but the total number of bonds in the lattice.
Interestingly, this $\tilde{B}\left( p,z\right) $ contains an extra term ($%
\tilde{B}\left( 0,z\right) $) which carries the information on the bonds 
{\em never }traversed. So, $d\left[ L^d/\left( 1-z\right) -\tilde{B}\left(
0,z\right) \right] $ is just $\sum_{t=0}^\infty z^tB(t)$. For completeness,
we report the result for unidirectional traverses: 
\[
\tilde{B}\left( p,z\right) =\frac{L^d}{1-z}-\frac{z\left( 1-p\right) }{%
\left( 1-z\right) ^2\left[ 2d+\left( 1-p\right) zG\left( z\right) \right] }.
\]
Finally, note that finite size effects are fully incorporated in this
approach, although they are implicitly ``buried'' in $G\left( z\right)
=L^{-d}\sum_{\left\{ k_i\right\} }\left[ 1-\left( z/d\right)
\sum_{i=1}^d\cos k_i\right] ^{-1}$ (where the sum is over the set $\left\{
k_i\right\} $ of allowed $L^d$ wavevectors). Needless to say,
generalizations to strip- or slab-like samples ($L_1\leq L_2\leq ...$) are
straightforward. Thus, it is possible to study the crossover of our
distributions, at least in principle, when the polymer length exceeds the
shortest dimension significantly ($t\gg L_1$). Physically, ultra-thin
membranes made from extra-long polymers can be manufactured. Until they
become reality, however, it may not be worthwhile to extend our analysis to
this class of crossover behavior.

A second remark concerns the
statistics of multiple {\it visits to sites}, as opposed to traversals
of bonds. Visits have 
been of considerable interest in the literature (see Hughes 
\cite{Hughes} and references cited therein). The average number $V_m(t)$ of
sites visited exactly $m$ times by a $t$ step random walk was studied by
Montroll and Weiss \cite{MontrollWeiss} and by Barber and Ninham \cite
{BarberNinham}. The $t\to\infty$ limit at fixed $m$ was considered by Hughes 
\cite{Hughes} (see also \cite{Newman}) in $d=1$ and by Montroll and Weiss 
\cite{MontrollWeiss} (see also \cite{ErdosTaylor}) in $d=2$; however, the
scaling limit expression of $V_m(t)$ has not to our knowledge appeared in
the literature. Since the analysis of $V_m(t)$ runs exactly parallel to that
of $B_m(t)$, we content ourselves to state the results here.

In dimension $d=1$ the average number of sites visited $m$ times is, to
leading order in the scaling limit, equal to the average number of bonds
visited $m$ times: $V_m(t)=2\,\mbox{erfc}(\mu_1)$\, for $t\to\infty$ at $%
\mu_1$ fixed. In dimension $d=2$ one has in terms of the scaling variable $%
\mu^{\prime}_2=\pi m/\log 8t$ 
\begin{equation}
V_m(t) = \frac{\pi^2 t}{\log^2 8t}\,\ee^{-\mu^{\prime}_2}\,\Big[\,1\,+\,
(\mu^{\prime}_2-2)\frac{\frac{\pi}{2}-1+C}{\log 8t}\,+\,{\cal O}%
(\log^{-2}8t) \Big]  \label{resultV2}
\end{equation}
valid for $t\to\infty$ at $\mu^{\prime}_2$ fixed. Comparison of Eqs.\,(\ref
{resultV2}) and (\ref{final2}) shows that to leading order $V_m(t)$ and $%
B_m(t)$ are identical up to a coefficient and a scale factor; however, the
coefficients of the first correction terms are different. The leading order
relation between $V_m(t)$ and $B_m(t)$ in $d=1,2$ is heuristically clear as
follows. Given a large number $m$ of visits to an arbitrary site $x$, there
will have been typically $m/(2d)$ traversals, starting from $x$, of a
specific bond $(x,x+\delta)$. Now the same bond will have been traversed,
typically, the same number of times in the opposite direction. So for each
site visited $m$ times, there are $d$ bonds traversed $m/d$ times.

Although we have not shown so explicitly, the scaling function in $d=2$ is
expected to be universal, {\it i.e.} lattice structure independent. In
dimensions $d>2$ the random walk is transient and it is easy to show that as
a consequence in the large $t$ limit $B_m(t)\simeq b_mt$\thinspace and
\thinspace $V_m(t)\simeq v_mt\,$, where $b_m$ and $v_m$ are nonuniversal.
For random walks that are not simple ({\it i.e.} have a step size
distribution not limited to nearest neighbor steps), bond traversals are not
unambiguously defined. However, visits to sites still are, and for such
walks $V_m(t)$ is readily calculated by the present method. One case of
interest is the scaling limit of $V_m(t)$ for lattice L\'{e}vy flights (also
called Riemann walks) \cite{GillisWeiss,Marizetal}, for which the
distribution of step sizes decreases as a power law.

\section*{Acknowledgments}

We thank J. Das, Manoj Gopalakrishnan, B. Schmittmann, and U.C. Tauber for
helpful discussions. 
T.A. and H.J.H. acknowledge the warm hospitality of the Physics
Department of Virginia Tech, where part of this research was conducted. 
This research was supported in part by grants from 
the Swiss National Science Foundation, the Hungarian Academy of Sciences
(Grant No.\ OTKA T029792),
and the US National Science Foundation through the Division of Materials
Research. The Laboratoire de Physique Th\'eorique in Orsay is associated
with Centre National de la Recherche Scientifique 
as research unit UMR8627.

\end{document}